\documentclass[draft]{agujournal}
\journalname{Geophysical Research Letters}

\begin{document}


\title{Migrating Scarps as a Significant Driver for Cometary Surface Evolution}

\authors{S.P.D. Birch\affil{1}, A.G. Hayes\affil{1}, O.M. Umurhan\affil{2}, Y. Tang\affil{1}, J-B. Vincent\affil{3}, N. Oklay\affil{3}, D. Bodewits\affil{4}, B. Davidsson\affil{5}, R. Marschall\affil{6}, J.M. Soderblom\affil{7}, J.M. Moore\affil{2}, 
P.M. Corlies\affil{1}, and S.W. Squyres\affil{1}}

\affiliation{1}{Cornell University, Ithaca NY, USA}
\affiliation{2}{NASA Ames, Moffett Field CA, USA}
\affiliation{3}{Deutsches Zentrum f{\"u}r Luft -und Raumfahrt (DLR), Institut f{\"u}r Planetenforschung, Asteroiden und Kometen, Berlin, Germany}
\affiliation{4}{Physics Department, Auburn University, Auburn, AL, USA}
\affiliation{5}{Jet Propulsion Laboratory, Pasadena, CA, USA}
\affiliation{6}{International Space Science Institute (ISSI), Bern, Switzerland}
\affiliation{7}{Department of Earth, Atmospheric and Planetary Sciences, Massachusetts Institute of Technology, Cambridge, MA, USA}

\correspondingauthor{Samuel Birch}{sb2222@cornell.edu}

\begin{keypoints}
\item We report the discovery of transient, migrating depressions within the smooth terrains on comet 67P
\item Migration is via uniform scarp retreat, which removes regolith completely from the nucleus and is driven by sublimating subsurface water ice
\item We develop a model that fully captures our own, and many previous, disparate observations, with implications for how comets seasonally erode
\end{keypoints}
%

%
%
\begin{abstract}
Rosetta observations of 67P/Churyumov-Gerasimenko (67P) reveal that most changes occur in the fallback-generated smooth terrains, vast deposits of granular material blanketing the comet's northern hemisphere. These changes express themselves both morphologically and spectrally across the nucleus, yet we lack a model that describes their formation and evolution. Here we present a self-consistent model that thoroughly explains the activity and mass loss from Hapi's smooth terrains. Our model predicts the removal of dust via re-radiated solar insolation localized within depression scarps that are substantially more ice-rich than previously expected. We couple our model with numerous Rosetta observations to thoroughly capture the seasonal erosion of Hapi's smooth terrains, where local scarp retreat gradually removes the uppermost dusty mantle. As sublimation-regolith interactions occur on rocky planets, comets, icy moons and KBOs, our coupled model and observations provide a foundation for future understanding of the myriad of sublimation-carved worlds. 
\end{abstract}

\section{Introduction}
Comets are a fundamental Solar System building block, yet many of the processes driving mass loss and the evolution of their bulk composition remain poorly understood, despite decades of observations. Early Rosetta observations of comet 67P/Churyumov-Gerasimenko (67P) led to hypotheses that sublimation-driven erosion of the consolidated nucleus is a principal erosional process \citep{Vincent2016}. The remnants of this process are the smooth terrains: vast deposits of fallback material of centimeter/decimeter-sized particles \citep{Thomas2015,Keller2017} that blanket large portions of the 67P's northern hemisphere, but are very sparse in the south \citep{Birch2017}. The formation of these smooth terrains is relatively well understood, where particles liberated by the intense insolation at perihelion of the southern hemisphere follow ballistic trajectories to cold, gravitational lows in the northern hemisphere \citep{Keller2017}. Most of this mass transport occurs seasonally during perihelion and is primarily driven by 67P's obliquity \citep{Keller2017}. 

The ubiquity of smooth terrains on other observed comet nuclei suggests that they play a fundamental role in shaping the surfaces of comets \citep{Cheng2013,Sunshine2016}. Comet 103P/Hartley 2 has a distinctive `waist' of smooth terrain \citep{Thomas2013a} that has been demonstrated to be the terminus for the deposition of airfalling particles ejected from the tip of the comet's nucleus \citep{Steckloff2016}. Similarly, 9P/Tempel 1 has large regions of smooth terrain material  \citep{Ahearn2011,Veverka2013} that retreated in between the Deep Impact and Stardust/NeXT flybys \citep{Thomas2013b}. Finally, though imaged with relatively poorer resolution, 19P/Borrelly has a similar waist to Hartley 2 \citep{Soderblom2002,Britt2004}, suggesting a similar formation. Like on 67P, observations of these nuclei led to hypotheses that the erosion of consolidated material dominated the mass loss on comets, and that the smooth terrains were principally lag material unable to escape during the ejection process \citep{Britt2004,Sunshine2016}.

Few changes of the consolidated material were observed during the Rosetta mission \citep{Pajola2017}, indicating that this process acts on multi-orbit timescales. Instead, many rapid changes were detected in what were previously thought to be the comparably inactive smooth terrains. These changes occurred in the the months leading up to perihelion, and include both morphological changes like depressions \citep{Groussin2015,ElMaarry2017} and `honeycombs,' \citep{Hu2017}, and spectroscopic \citep{Fornasier2016,DeSanctis2015}. Together, these changes suggest the nucleus removed an overlaying layer of dust prior to receiving its newest layer at perihelion. Yet to date, no satisfactory explanation has been offered to describe this seasonal mass loss, and so our understanding of the most basic processes capable of modifying cometary surfaces remains limited. 

\section{Discovery and Migration of Depression Scarps}
Using image data from the Optical, Spectroscopic, and Infrared Remote Imaging System \citep{Keller2007} (OSIRIS) Narrow angle camera (NAC) onboard the Rosetta spacecraft, we find a cluster of decameter-scale arcuate depressions (hereafter called depressions), with prominent scarps that form their upper boundaries (with respect to the individual georeferenced images; Figure 1). These features are located in the Hapi region \citep{ElMaarry2015} within the comet's `neck' (see Figure S1 for the regional geometry), a gravitational low that occupies $\sim$4\% of the comet's surface, where the largest volumes of fallback materials are predicted to be deposited each perihelion passage \citep{Thomas2015}. 

\begin{figure}
\includegraphics[scale=0.3]{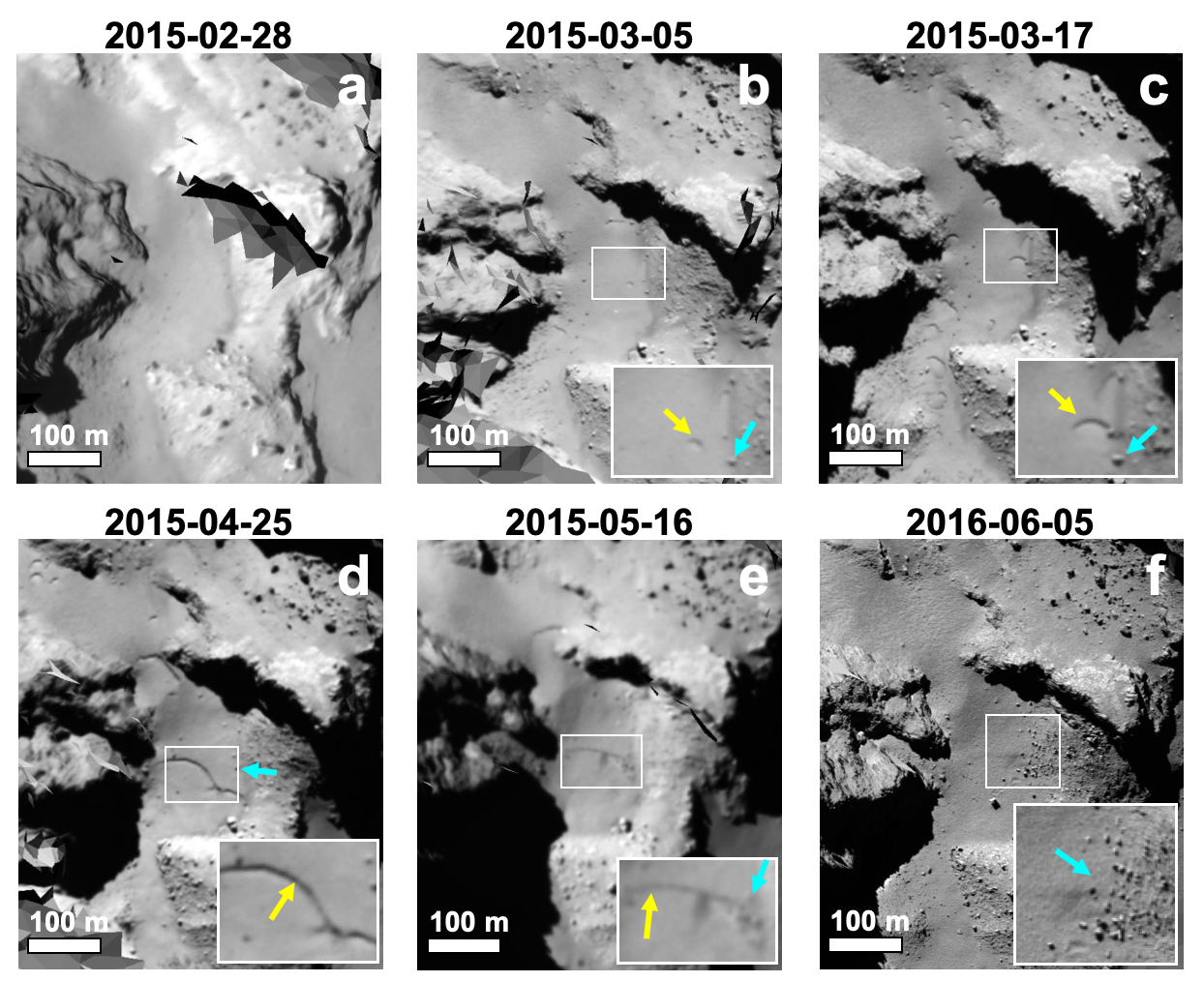}
\caption{(Georeferenced images of the Hapi region. (a) On February 28, 2015 the smooth plains are uniform and featureless. Depressions (yellow arrows in all panels) began to appear in this region in March 2015 (b/c), and then coalesced into larger features over a 2+ month period (d/e). Images 12 months later (after perihelion; panel f) show a similar surface prior to depression formation (a), with featureless smooth terrains. Underlying boulders, however, have been excavated (cyan arrows) as a result of the changes in panels b-e. Note that artifacts in the images are due to image projections, and they do not alter our interpretations.}
\label{fig1:pitevolution}
\end{figure}
The depressions first appear on March 5, 2015 as small arcuate features proximal to boulders or cliffs (yellow arrows in Figure 1b). They then enlarge and migrate 11 meters over the subsequent 12 days (equivalent to an average rate of 3.8$\pm$0.6 cm/hour) into quasi-circular, decameter-sized depressions (Figure 1c). The migrating boundaries of each depression are all scarps (Figure 2) that originate from the initial arcuate features (shown in Figure 1b).

Additional images of this region were acquired over the following $\sim$2.5 months (Figure S2/Table S1 in the \textit{Supplementary Material (online)}, hereafter `SOM'); the final high-resolution image of the depressions in this region was acquired on May 22, 2015. This sequence of observations clearly shows that the initial depressions (Figure 1c) evolved into fewer, larger features that span nearly the entire plain (Figure 1d/e). The scarp migration rate at these later dates also increases slightly to 6.1$\pm$1.5 cm/hour (31 meters in 21 days), a change that is noticeable in the registered images (Figure S2). Images acquired throughout this time period also show dust jets emanating from the region (Figure S4), suggesting that, as the depressions form, most material is removed directly to the coma (see SOM). 

The curvature of the depression scarps is consistent with growth and agglomeration of numerous small depressions into increasingly large depressions via uniform retreat of the depression scarps \citep{Howard1995,Hayes2017}. These observations suggest that an upper layer of regolith is progressively removed in the months just prior to perihelion (see SOM), leaving behind only the largest boulders (cyan arrows in Figure 1e/f; Figure S3). We rule out mass wasting as the origin for these boulders, as we do not observe a corresponding change in the upslope cliff \citep{Pajola2017}. Post-perihelion observations show a near-reversion to the original featureless surface, which we attribute, in part, to deposition during polar night \citep{Keller2017}. 

For all depression scarps, the migration direction is toward the top of each panel in Figure 1, a direction that is not aligned with the Sun. Instead, migration is nearly opposite the Hathor cliff, towards the comet's large lobe (see Figure S1 in the SOM). This implies that the depression scarps are migrating oppositely compared to all other known sublimation-formed depressions across the Solar System \citep{Rhodes1987,Moore1996,Thomas2000,Betterton2001,Moore2017,Howard2017}, which all enlarge along a direction near-parallel to the solar azimuth, for non-nadir solar elevations.

\section{Topographic Evolution of Depression Scarps} 
We measure the depression's topography using photoclinometry \citep{Kirk2003}. Our method is specifically tailored to the smooth terrains of 67P \citep{Tang2018}, and allows us to precisely measure the evolution of the depth of the depressions (Figure 2). This method, unlike the global stereophotoclinometry shape model (SPC \citep{Jorda2016}), uses a single OSIRIS image to quantify the topography at the scale of the image resolution \citep{Birch2017,Tang2018}. Over short distances (a subset of a given OSIRIS NAC image), we can improve the vertical resolution, as compared to the global shape model, by an order-of-magnitude (see SOM). Over longer distances, errors can accumulate, and so we restrict ourselves to regions $<$500$\times$500 m$^2$ (see SOM). 

For each image in which changes are detected and sufficiently well resolved (Figure 1c/d), we divide the image into multiple subsections (of which there are four total) that include the changes and only a limited amount of the surrounding terrains. This ensures that we do not introduce errors that may result from photometric variations across larger regions and/or terrain types (see SOM). We then input each of those subsets into our DTM-generation pipeline \citep{Tang2018}. This process outputs a three-dimensional surface of the nucleus, which we then manually inspect for any artifacts. Finally, for each depression in each of the DTMs, we record the maximum depth of the depression, relative to the surrounding terrains, along with the vertical accuracy of the DTM.

\begin{figure}
\includegraphics[scale=0.47]{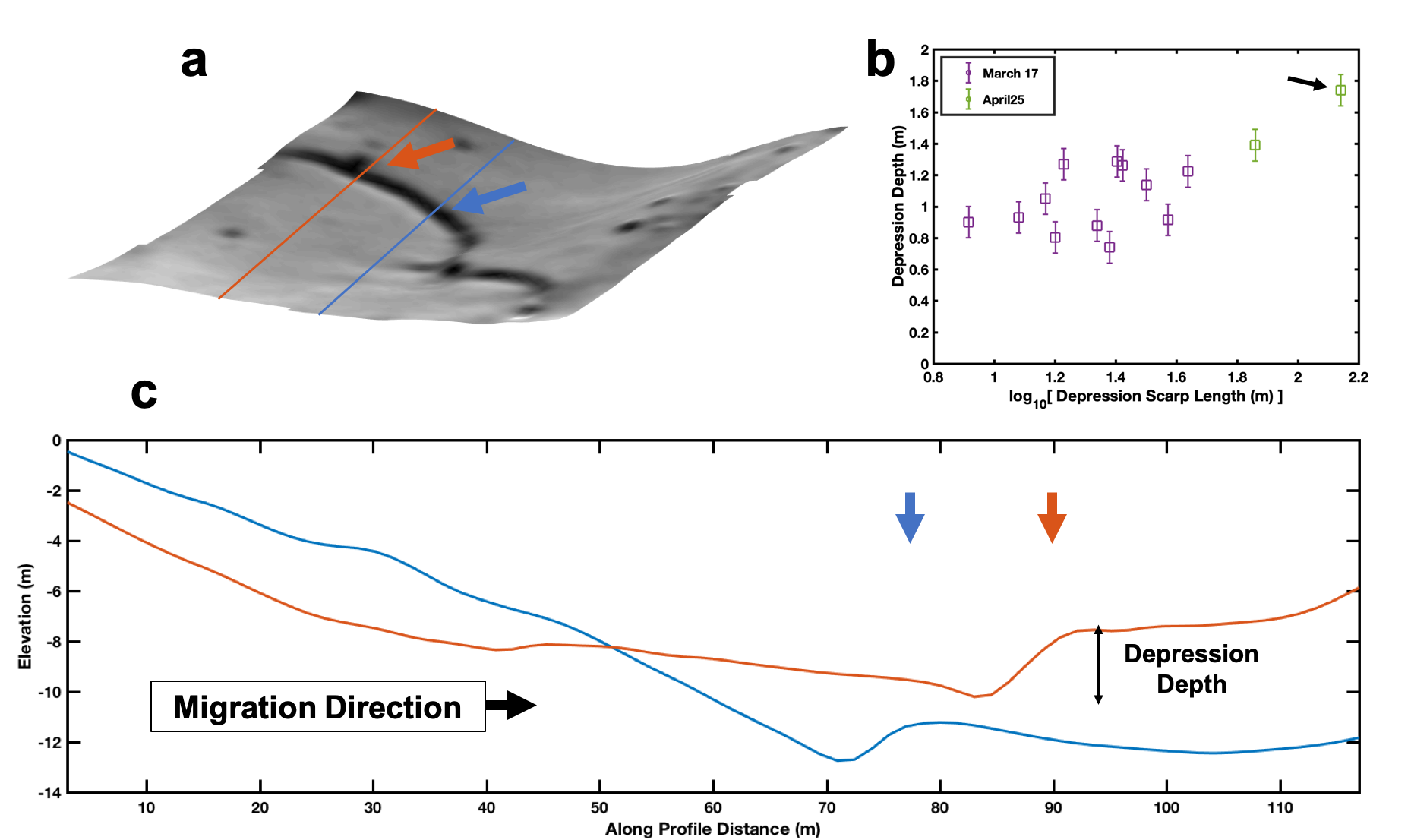}
\caption{(a) An example DTM of the depressions as imaged on April 25; (b) Depression depth as a function of the measured scarp length for March 17 (purple) and April 25 (green). The black arrow denotes the point measured from the DTM in panel a; (c) Two profiles of the depression scarp on April 25, where the color of the lines corresponds to the line profiles in panel a. The arrows denote the location of the depression scarp. Depression depths are measured as the difference between the bottom of the depression and the immediate exterior, after de-trending for the long wavelength slope of the region.}
\label{fig2:topography}
\end{figure}
Unlike other depressions formed by sublimation erosion, such as suncups on Earth \citep{Rhodes1987,Moore1996,Betterton2001}, polar depressions on Mars \citep{Byrne2003}, and pits on Pluto \citep{Moore2017,Howard2017}, the depressions in Hapi are neither symmetric nor flat-floored. Instead, the depressions progressively deepen towards their scarps (Figure 2c). Since the DTM tilt is controlled (see SOM), it is also clear that the depressions form on slopes, with the depression scarps migrating downslope (Figure 2c). 

Photoclinometry also allows us to make DTMs for each image, providing information over time as the depressions evolved. We observe that larger/longer depressions excavate to greater depths (Figure 2b), implying that the lower boundary of the Hapi smooth terrain basin, which would potentially limit further excavation, is much deeper than the depression depths (Figure S1). This is consistent with our post-perihelion observations of the region (Figure 1f), in which no underlying consolidated layer (e.g., such as the one inferred from Philae measurements \citep{Biele2015} at the Agilkia landing site) is observed. Instead, we only observe more smooth terrain materials and previously buried debris (Figure S3). Thus, the observed relation between excavation depth and size suggests the process only an entire upper $\sim$1 m (see SOM) of regolith materials. The only process capable of arresting the scarp growth is lateral expansion into the consolidated nucleus, which we assume occurs, followed by the possible deposition of a new layer after Hapi goes into polar night. 

\section{Observational Evidence of Subsurface Water Ice and Mass Loss}
On April 25, the depressions were imaged with multiple OSIRIS NAC color filters, allowing for a spectrophotometric analysis. Through a decrease of the spectral slope (i.e., a blueing of the surface \citep{Oklay2015,Oklay2016,Barucci2016,Pajola2017}), we identify exposures of water-ice (Figure 3). The use of these data, however, requires various corrections. We first perform a re-alignment of the images to a common datum (which we choose to be the 701.2-nm image) to sub-pixel accuracy to remove any color artifacts that may arise due to image misalignments. This has been done for similar analyses (see \citet{Oklay2016}). To eliminate effects resulting from varying illumination conditions, a Lommel-Seelinger disk correction is applied \citep{Hapke2002} to each image, assuming the phase function and phase reddening parameter that had been measured for the bulk nucleus \citep{Fornasier2015} at 649.2 nm. Photometric angles are calculated using latest SPICE kernels and the 3D shape model by \citet{Preusker2017}. Finally, to increase the signal-to-noise ratio, for each image, we spatially bin the data (as denoted by the colored boxes in Figure 3b).

The spectral slope ($S$) that is shown in Figure 3c/d is then defined as: 
\begin{equation}
S(\%/100 nm) = \frac{(R_{882.1} - R_{480.7})\times10^4}{R_{480.7} \times (882.1 - 480.7)},
\label{spectral_slope}
\end{equation}
where $R_{882.1}$ and $R_{480.7}$ refer to the radiance factor (I/F) as measured at 882.1 nm and 480.7 nm, respectively. Spectral slopes ($S$) were calculated for each non-shadowed pixel (of which there are few; Figure S7); any `blueing' of the surface indicates a lower spectral slope than the bulk nucleus (Figure 3d; typically $\sim$18\%/100 nm \citep{Fornasier2015,Oklay2016}). Note that there were very few pixels where we did not get a significant signal (Figures S7/S8) as it is possible to observe into shadowed areas due to the high dynamic range of the OSIRIS cameras (see SOM). Thus, as in previous studies we assume that a similar `blueing' of the surface ($S$$\leq$14-15\%/100 nm) indicates water-ice exposures and/or activity \citep{Vincent2015}. Furthermore, not only is use of the spectral slope a reliable method \citep{Oklay2015}, but it also ensures consistency between our results and previous investigations that studied similar regions hosting activity across the nucleus \citep{Pommerol2015,Oklay2016,Fornasier2016,Barucci2016,Pajola2017,Laforgia2015,Lucchetti2016,Fornasier2015,Vincent2015}. Our choice of the 882.1 nm and 480.7 nm filters was out of necessity, as they were the only common pairs between all 11 images, though our choice does not affect our fitting \citep{Oklay2015}. 

For the depressions where water ice was present on April 25, we chose four study locations (that each comprise from 11-25 pixels; Figure 3b), two within each of the observed scarps, and two on the immediate surrounding smooth terrains (Figure 3b). We also analyzed two similarly-sized areas on distant smooth terrains of 67P for reference (yellow arrows in Figure 3a), as they are assumed to be representative of average (Non-Hapi) smooth terrain surfaces. For each region, we extract the mean I/F from each of the 11 images and then normalize these values relative to the I/F measured at 480.7 nm. These relative reflectances are plotted in Figure 3c. 
\begin{figure}
\includegraphics[scale=0.67]{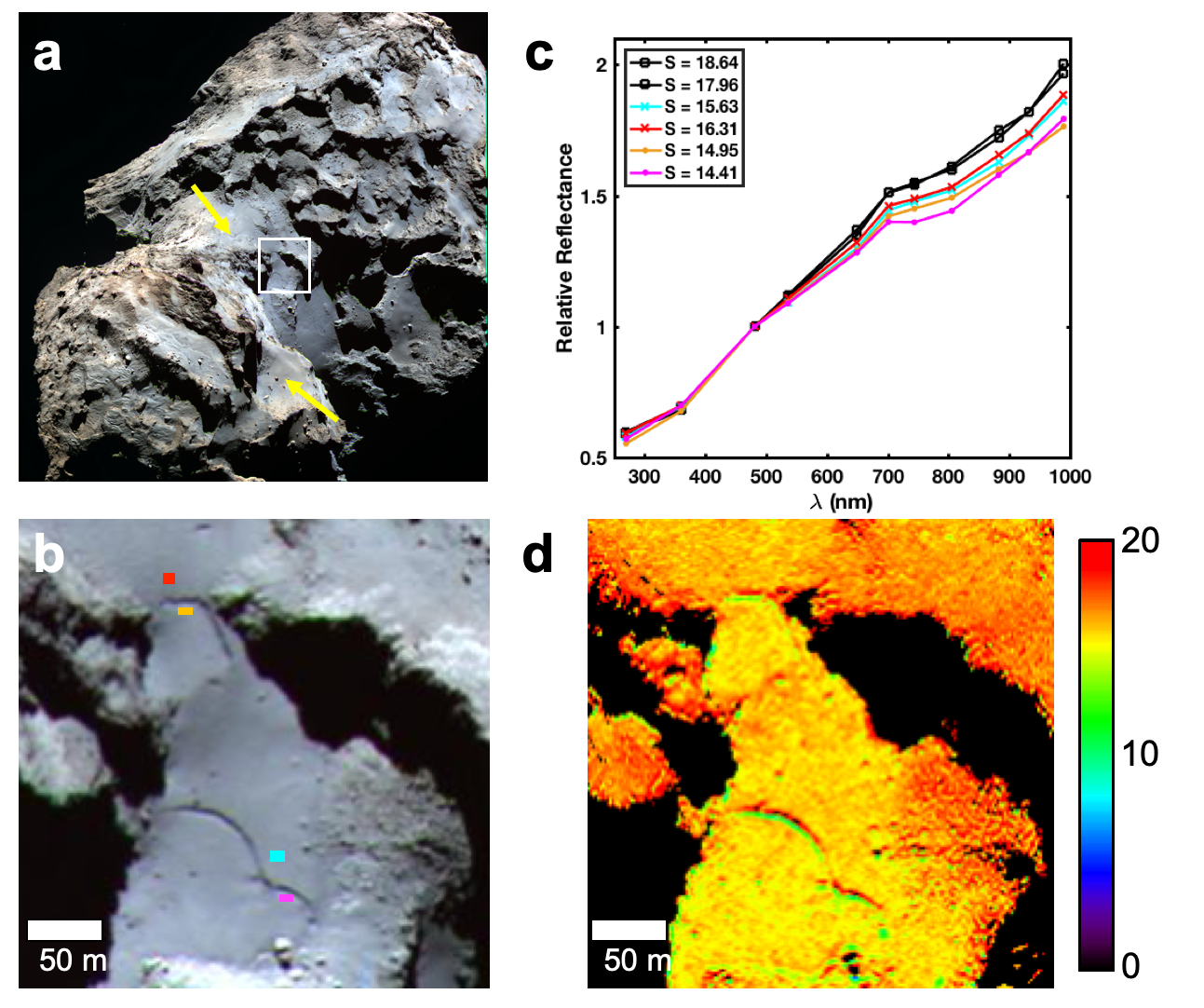}
\caption{(a) High resolution RGB OSIRIS NAC image from April 25; (b) RGB image of the region within the white box in panel a. The colored squares are regions considered in the analysis in panel c; (c) Mean relative reflectance spectra of the four regions in panel b, along with the reflectance spectra for the regions marked by yellow arrows in panel a (black lines). The spectral slope $S$ for each region is given in the top left. Errors on the I/F values are smaller than the data points; (d) Spectrophotometric image of the region, where the color of each pixel denotes the spectral slope value.}
\label{fig3:color}
\end{figure}

This analysis clearly shows three clusters of materials: Non-Hapi smooth terrain surfaces (square points in Figure 3c), Hapi-like \citep{Fornasier2015} (crosses in Figure 3c), and icy material (dots in Figure 3c). As expected, the Non-Hapi surfaces have the highest spectral slope, similar to previous studies \citep{Oklay2016}. This is consistent with dehydrated material at depths less than the thermal skin depth. The Hapi-like materials have slightly lower spectral slopes, and have been shown to be consistently more ice-rich than other regions on 67P \citep{Oklay2016}. The most ice-rich materials, with the lowest spectral slopes, are located within the scarps, and in other distant regions, proximal to the local terminator. The former are consistent with an exposure of subsurface ice because, where the depressions are located, it has been $>$2 hours after local sunrise (Table S1). The latter lie along the local terminator (i.e. sunrise) and are consistent with a morning surface frost \citep{DeSanctis2015,Fornasier2016}.

We also observe a direct link between 67P's coma and the depressions \citep{Shi2018}. OSIRIS observations from the same period (March 14 \& May 10) show a significant amount of dust coming from the location of the depressions (see SOM; Figure S4/S10). This is consistent with our model (Section \ref{model}), and provides a direct link between mass loss processes and observable large-scale surface changes.

\section{A Self-Consistent Model to Describe Smooth Terrain Erosion} \label{model}
In the months just prior to perihelion, the Hapi region was receiving more insolation than it does at any other point in 67P's orbit \citep{Keller2017}. Based on the observed formation and evolution of these features, as well as the presence of dust jets, the likely geomorphic process driving the depression's scarp retreat is the sublimation of near-surface ices. This seasonal erosion occurs only for this few-month timespan before perihelion, with deposition from the southern hemisphere likely dominating the surface evolution during and after perihelion \citep{Keller2017}. To model the migration and formation of these features during this short, erosion-dominated timespan, we require some simplifying, though well-justified, assumptions.

We assume that the regolith in Hapi is similar to the material at Agilkia, Philae's first touch down site, consisting of centimeter/decimeter-sized gravels (see SOM). That the Hapi smooth terrains are far deeper than the measured depression depths requires that sublimation of ices must come from within the regolith layer itself, and not from an underlying consolidated layer. This implies that dust and ices are well mixed, so that exposed water-ice freely sublimates. This ice is assumed to be co-deposited with the particles following their ejection from the southern hemisphere \citep{Thomas2015}, and is different from the surficial daily frosts \citep{DeSanctis2015}. We then assume that the surrounding surfaces are comparably inactive \citep{Shi2016}, and calculate their temperature throughout the period where changes are occurring (see SOM)). 

Growth and migration of the depressions are then driven by re-radiated solar insolation from the depression floors and surrounding terrains onto the water ice-enriched depression scarps (Figure 4a \& Figure S1). The total power flux of energy ($F$) emitted from the depression floors and surrounding terrains, onto the scarp is:
\begin{equation}
F = \epsilon \sigma T_s^4 + \frac{S_{\odot}A}{{r_h}^2},
\label{flux}
\end{equation}
where $\epsilon$ is the emissivity, $T_s$ the surface temperature, $A$ the albedo, and $r_h$ the heliocentric distance. The first term describes the flux re-radiated energy and the second is the flux from reflected light. However, water ice absorbs more efficiently in the infrared and the albedo of the surface is very low, implying that contributions from reflected radiation will be minimal (see SOM). Thus, though the fluxes from re-radiated and reflected light are of the same order-of-magnitude, we only consider the effects from the re-radiated light. We ignore directed insolation as the scarps are never fully illuminated by the Sun (Figure S6).
\begin{figure}
\includegraphics[scale=0.5]{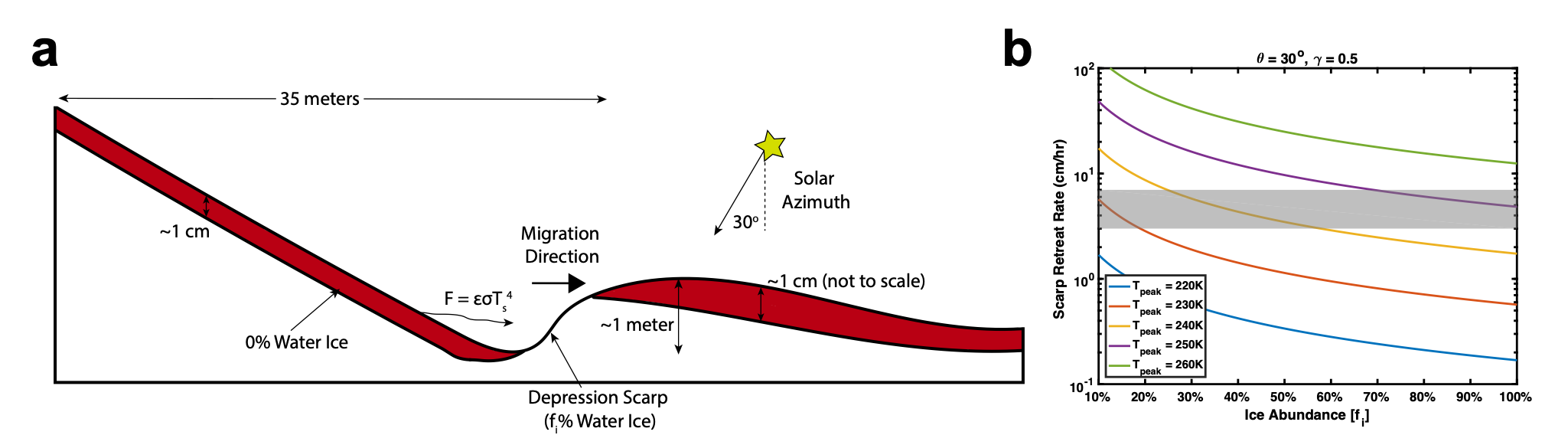}
\caption{(a) Conceptual model, with the profile extracted from Figure 2c. The migration direction of the depression scarp is toward the right, while the solar azimuth is pictorial shown for one portion of the day. The red upper layer illustrates the comparably inactive top layer that has a thickness on the order of the thermal skin depth \citep{Shi2016}; (b) Depression scarp retreat rates are calculated for peak temperatures ($T_{peak}$) according to Equation \ref{main}. The measured scarp retreat rates in all images are plotted in the grey box.}
\label{fig4:model}
\end{figure}

We assess sublimation rates of the ice within the granular matrix of the depression scarp using the formulation developed by \citet{lebofsky}, supplemented by the recent work of \citet{Hobley2018}, using models for both the peak surface temperature and temperature variations throughout a given day (see SOM). We define the {\it{daily-averaged}} flux of sublimated water-ice from the exposed scarp as:
\begin{equation}
q  = \gamma\frac{P_{vap}(T_{peak})}{2\pi \sqrt{L}}.
\label{lebofsky}
\end{equation}
Equation \ref{lebofsky} was applied recently by \citet{Hobley2018} to model ablation rates of water-ice on the surface of Europa, and we use it here to estimate fluxes of sublimated water-ice. Equation \ref{lebofsky} describes the sublimation flux, $q$, of volatile ice from within the exposed subsurface. This quantity is dependent upon the latent heat of sublimation of water ice, $L \approx 2.85\times 10^3$kJ/kg and its vapor pressure
\begin{equation}
 P_{vap} = 3.56\cdot 10^{12}\exp\left(-\frac{6141.667}{T_{s}}\right),
\label{vaporpressure}
\end{equation}
evaluated at the peak daily temperature $T_{peak}$. Note that most of the scarp retreat happens at times near this peak temperature owing to the Arrhenius form of the vapor pressure. 

The quantity $q$ is derived based on the assumption of a daily sinusoidal temperature dependence $-$ an appropriate representation for standard spherically symmetric bodies. For such bodies, the peak temperature coincides to local noon. On more complex-shaped bodies like 67P, $T_{peak}$ may occur at times not coinciding with local noon. For instance, during the March$-$May time frame, $T_{peak}$ on Hapi occurred close to local sunrise (Figure S5). Under such situations, Equation \ref{lebofsky} will overestimate the amount of sublimated gas. We include a factor $0<\gamma<1$ that takes into account the effective decrease of sublimated gas. In the particular case of Hapi where the small lobe delays local sunrise (see SOM), the shape of the predicted temperature profile (Figure S5) over the course of one day shows a strong increase near sunrise and then diminishes in a sinusoidal fashion beyond it. The shape near $T_{peak}$ appears locally like half of a sinusoid and, as such, we adopt a value $\gamma = 1/2$.

The actual rate of migration of the depression's scarp, normal to the exposed surface, $\dot z_n$ is given by the expression 
\begin{equation}
\dot z_n = q/\rho_{i},
\end{equation}
which is dependent on the effective density of water ice, $\rho_{i}$, an unknown quantity which we parametrize. 

When the water ice in the regolith has sublimated, the gas emerges in a direction normal to the exposed scarp surface. The liberated gas moves with its typical Maxwell-Boltzmann speed $v_a = c_s/\sqrt{2\pi}$ where $c_s \equiv \sqrt{kT_s/m_{H_2O}}$, which is a function of the temperature, the Boltzmann constant $k$, and the mass of a water molecule $m_{H_2O}$. The density of the gas upon emergence can be estimated to be $\rho_{gas} \approx q/v_a$. This expanding vapor carries the remaining non-ice material away from the depression, with only the largest regolith particles ($>$1 m), at low temperatures ($T_{peak}<$ 235 K) remaining in the depression (Figure S8/S9). The remainder of the material likely escapes the region completely (see SOM).

Given 67P's mean bulk density $\rho = 538$ kg/m$^3$ \citep{Patzold2019}, we express the uncertainty in the relative content of subsurface water ice by the parameter $f_i$, whereby $\rho_{i} = f_i \rho$. To predict the actual observed propagation rate of the sublimation front parallel to the local surface, which we call $\dot z_{surf}$, we must take into account the geometric projection that arises due to the angle formed between the local surface normal of the exposed scarp and the surface normal of the surrounding landscape. In this case $\dot z_{surf} = \dot z_{n}/\sin\theta$ where $\theta$ is the angle formed between the mean surface normal vector and the normal vector of the exposed scarp. Based on the local DTM of the depression scarps (Figure 2c) we see that the scarp angle with respect to the local slope falls in the range $\theta \approx 20^{\circ} - 35^{\circ}$. The single expression for $\dot z_{surf}$ is therefore
\begin{equation}
\dot z_{surf} = \gamma\frac{P_{vap}(T_{peak})}{2\pi \sqrt{L} f_i \rho \sin\theta}.
\label{main}
\end{equation}
The above expression for the observed depression scarp migration rate does not take into account the situation when the ice fraction $f_i$ gets small enough that the propagation of the sublimation front gets quenched by the larger effective volume of non-ice debris. While we do not intend to model that process here, we nominally assume that once $f_i < 0.1$ then no propagation takes place. Equation \ref{main} is calculated for varying subsurface, volumetric ice abundances $f_i$ (Figure 4b), and it is clear that migration is consistent for temperatures $\ge$230 K (close to our modeled probable temperatures, see SOM). However, we require an enhancement in the volumetric ice fraction in the scarp compared to the surrounding surface, between 20\% and 50\%, to match the observed migration rates (Figure 4b). These values are consistent with the global dust/ice ratio as measured by GIADA \citep{Rotundi2015}. 

Finally, we assume that the same process drives the initial formation of the depressions, with one key difference: the radiation that drives sublimation is from the boulders or cliffs to which all the depressions are initially proximal. Thus, the boulders and cliffs act as seed points for an instability that propagates across the entire plain. 

\section{Implications for the Evolution of Cometary Surfaces}
Our combined model and OSIRIS observations show that material is lost from Hapi's smooth terrains not by a gradual diffusive process, but by localized bursts of sublimation within migrating depression scarps. For just a short time ($\sim$2.5 months), these scarps grow and migrate across the smooth terrains, removing a layer of smooth terrains from the nucleus entirely (see Figure S10). It is not unreasonable to expect that other smooth terrains on 67P lose mass in a similar way. In fact, our combined observations and model tie back to numerous past observations made by the Rosetta mission, and provide a critical missing link as to how 67P's smooth terrains seasonally erode.

For example, while \citet{Keller2015} described that self-heating within 67P's neck could explain the observed early activity, our model provides the missing link as to how material is ultimately removed. Similarly, migrating scarps of smooth terrains were observed by \citet{ElMaarry2017} in the Anubis region and \citet{Groussin2015} in Imhotep though with faster migration rates ($\sim$ 22.5 cm/hour and $\sim$18-28 cm/hour respectively). As the observed process appears to be the same between these changes - migrating scarps with ice embedded within them \citep{Groussin2015} - our model can readily describe the evolution of these changes, albeit likely with higher surface temperatures. Finally, indirect detections of water outgassing from smooth terrains \citep{Marschall2017}, spectrophotometric variations of the nucleus near perihelion \citep{Fornasier2016,Ciarniello2016}, and observations of the numerous isolated bright patches on other smooth terrains in the months prior to perihelion \citep{Deshapriya2018} can all be related to bursts of migration from exposed ice within scarps of the smooth terrains.

Our model not only ties together our many disparate observations, but it also provides a consistent mechanism that can describe how 67P's, and cometary smooth terrains in general, seasonally shed dust. Similar smooth terrains are seen on all other resolved comets like Tempel 1, Hartley 2, Wild 2, and Borrelly as well \citep{Sunshine2016}, and so our work has considerable applicability. Due to the generality of our model and its derivation from first-principles, our work can be applied to other volatile-rich worlds such as ice-rich asteroids, KBOs, and icy airless moons. Comet 67P provides a window into the erosional processes responsible for shaping not only cometary surfaces but sublimation-dominated planetary surfaces in general.
%
%
\section{Acknowledgments}
This research was supported by Cornell University. We thank Mathieu Choukroun and an anonymous reviewer for providing helpful comments that improved the manuscript. We would also like to acknowledge the Principal Investigator of the OSIRIS camera on ESA's Rosetta spacecraft, Holger Sierks, and the ESA Planetary Science Archive for the data used in this study. This research has made use of the scientific software shapeViewer (www.comet-toolbox.com). Part of this work was carried out at the Jet Propulsion Laboratory, California Institute of Technology, under contract with NASA. Finally, we thank Nicholas Kutsop for assistance in preparing the manuscript. All OSIRIS image data presented in this paper were downloaded, and are freely available on ESA's Planetary Science Archive (https://archives.esac.esa.int/psa) and NASA's Planetary Science Data System (https://pds.nasa.gov). Topographic and spectrophotometric data that support the plots within this paper and other findings of this study are available from the corresponding author upon request. Additional data are also provided in the Supplementary Information. 

\bibliographystyle{plainnat}

\end{document}